\documentclass[aps,superscriptaddress,twocolumn,notitlepage]{revtex4-2}
\usepackage{times,color,amsfonts,amssymb,amsbsy,amsthm,amsmath,graphics,graphicx,hyperref,bm,bbm,makecell,mathtools,dsfont,upgreek,hyperref,fancyhdr,multirow,braket}
\usepackage[caption=false]{subfig}
\usepackage{tikz}
\usetikzlibrary{shapes.geometric}
\usetikzlibrary{decorations.pathmorphing}
\usetikzlibrary{shadings,3d}
\usetikzlibrary{snakes}
\usepackage{blindtext}
\usepackage[capitalise]{cleveref}
\usepackage{tcolorbox} 
\hypersetup{colorlinks,linkcolor={black},citecolor={black},urlcolor={black}}
\definecolor{psred}{rgb}{0.8,0.2,0.2}

\newcommand*{\arxiv}[1]{\href{https://arxiv.org/abs/#1}{arXiv: #1}.}
\newcommand*{\availableonline}[1]{\href{#1}{Available Online}.}

\newcommand*{\etal}[0]{\textit{et al.}}
 
\usepackage[utf8]{inputenc}
\usepackage[T1]{fontenc}

\usepackage{multirow}
\usepackage{makecell}
\usepackage{colortbl}

\usepackage[normalem]{ulem}

\ifx\showcorrections\undefined
    \newcommand{\cdel}[1]{}
    \newcommand{\cdeleq}[1]{}
    \newcommand{\cadd}[1]{#1}
\else
    \newcommand{\cdel}[1]{\textcolor{red}{\sout{#1}}}
    \newcommand{\cdeleq}[1]{\textcolor{red}{#1}}
    \newcommand{\cadd}[1]{\textcolor{blue}{#1}}
\fi

\begin{document}
\title{Quantum Illumination and Quantum Radar: A Brief Overview}
\author{Athena Karsa}
\affiliation{Department of Physics \& Astronomy, University College London, London WC1E 6BT, United Kingdom}
\affiliation{Korea Research Institute of Standards and Science, Daejeon 34113, Korea}
\affiliation{Department of Computer Science, University of York, York YO10 5GH, United Kingdom}

\author{Alasdair Fletcher}
\affiliation{Department of Computer Science, University of York, York YO10 5GH, United Kingdom}
\affiliation{nodeQ, The Catalyst, Baird Lane, York, YO10 5GA, United Kingdom}

\author{Gaetana Spedalieri}
\affiliation{Department of Computer Science, University of York, York YO10 5GH, United Kingdom}

\author{Stefano Pirandola}
\thanks{Corresponding author}
\email{stefano.pirandola@york.ac.uk}
\affiliation{Department of Computer Science, University of York, York YO10 5GH, United Kingdom}

\begin{abstract}
    Quantum illumination (QI) and quantum radar have emerged as potentially groundbreaking technologies, leveraging the principles of quantum mechanics to revolutionise the field of remote sensing and target detection. The protocol, particularly in the context of quantum radar, has been subject to a great deal of aspirational conjecture as well as criticism with respect to its realistic potential. In this review, we present a broad overview of the field of quantum target detection focusing on QI and its potential as an underlying scheme for a quantum radar operating at microwave frequencies. We provide context for the field by considering its historical development and fundamental principles. Our aim is to provide a balanced discussion on the state of theoretical and experimental progress towards realising a working QI-based quantum radar, and draw conclusions about its current outlook and future directions.
\end{abstract}

\maketitle

\section{Introduction}
\label{sec:intro}

The development of quantum information theory \cite{Nielsen:Book2000,braunstein2005quantum,weedbrook2012gaussian,serafiniBook} has led to a so-called `second quantum revolution' \cite{Dowling2003-2ndQRev} in developing new quantum technologies. By exploiting various quantum phenomena to achieve performances unattainable solely through classical means, vast technological progress has been made in a wide variety of fields including quantum cryptography~\cite{Pirandola2019-advQcrypt}, quantum computing~\cite{NatRevQComp,Fitzsimons:SQC2017} and quantum sensing~\cite{pirandola2018advances,Degen:SensingRev}. At the heart of many such advances lie fundamental properties such as the uncertainty principle and quantum entanglement~\cite{Horodecki:RevModPhys}; the latter represents the ability of quantum systems to form correlations so strong to go beyond any classical counterpart. 

Quantum illumination (QI)~\cite{lanzagorta2020opportunities,luong2020entanglement,daum2020quantum,shapiro2020story,torrome2023advances} is an example of such an emerging technology, capable of utilising entangled states to outperform corresponding classical benchmarks \cadd{in certain quantum sensing tasks}. \cadd{A remarkable aspect of QI is that the improved performance occurs despite the fact that all entanglement is lost in the process, an occurrence which, for almost all other quantum technologies, is completely detrimental.} Following preliminary work in 2005 demonstrating that entanglement can improve the distinguishability of entanglement-breaking channels~\cite{sacchi2005entanglement}, the QI protocol was later introduced in 2008~\cite{lloyd2008enhanced,tan2008quantum} using an entanglement-based approach to improve the ability to detect a weakly-reflecting object embedded in a bright thermal background. \cadd{The relevant pathway, or quantum channel, for the entangled states is characterised such that it is an entanglement-breaking one, whether or not the object is present.} The proposed technique remains highly important to the field and the general structure of the protocol has been often repeated. A source generates entangled pairs, only one member of which is used to probe the target region while the other is retained at the source awaiting recombination with the signal upon its eventual return. An optimal joint measurement of the pair is tasked with capturing information held by their entangled nature to yield improved sensitivity to the target detection problem. 

The initial models for QI, both with discrete-~\cite{lloyd2008enhanced} and continuous-variable~\cite{tan2008quantum} systems, were proposed in the optical regime, so as to provide a first extension of the LIDAR from the classical to the quantum realm. However, the more favourable regime for quantum advantage found by these earlier works was typical of non-optical frequencies (bright thermal noise). This led to the formulation of the QI protocol in the microwave regime in 2015~\cite{barzanjeh2015microwave}, where the background thermal noise is naturally stronger. By extending QI to the microwaves and showing quantum advantage, the 2015 work gave birth to the first theoretical prototype of quantum radar able to show quantum advantage over classical designs.

These initial QI results ignited a plethora of theoretical and experimental investigations with the goal of performing quantum target detection using QI. The desire for improved radar systems is ever-present for a wide range of applications, spanning the military to space exploration. \cdel{Moreover, s}\cadd{S}ince radar already utilises the reflection of electromagnetic radiation to to detect distant objects (targets) \cite{skolnik2008introduction}, the potential combination with \cdel{quantum illumination}\cadd{QI} to realise a quantum radar was a natural progression for research. 

However, whilst various QI protocols have been successfully demonstrated experimentally: from optical wavelengths~\cite{lopaevaexp,zhangent,zhangexp,england2019quantum} to first applications with microwaves~\cite{chang2019quantum,luong2019receiver,shabirQI,Huard2022} the prospect of a working quantum radar, \cdel{at least}\cadd{particularly} one based on entanglement, remains littered with many technological issues. Indeed\cdel{, although} there has been a great amount of fanfare within the defence community and particularly the media concerning military applications of quantum radars\cdel{,}\cadd{.} \cadd{Yet, }the prospects of a true quantum radar, capable of preserving quantum advantage over large distances in realistic applications\cadd{,} remain extremely challenging. Nonetheless, in the relatively short time that has passed since its inception, QI and our understanding of it have come a long way. While the attainment of the ultimate goal of a long-range quantum radar remains elusive, intermediate results do show that perhaps more modest applications of such a technology could certainly be a real possibility in the near future.

This review provides an overview of QI and its application to quantum target detection and the potential for a quantum radar. The aim is to cover the background and introduce some of the more recent advances to provide the reader with a general overview of the current state-of-the-art. We begin with a brief overview of the basic formulation of QI protocols in Sec.~\ref{sec:basic}. Then, in Sec.~\ref{sec:criteria}, the working criteria for assessing what constitutes quantum radar are outlined. Sec.~\ref{sec: theory} proceeds to discuss the relevant theoretical background of classical radar technology and hypothesis testing. QI is then presented in more detail covering applications from the optical to the microwave regime, and classical benchmarking is discussed. In Sec.~\ref{sec:exp reqs}, the theoretical and experimental challenges behind the implementation of QI as prototype of quantum radar are discussed. In Sec.~\ref{sec:QIexperiments} 
we provide a brief overview of the experimental progress. Finally, in Sec.~\ref{sec: outlook} we draw our conclusions.

\begin{figure*}[tb!]
\begin{flushleft}
\hspace{0.5cm} \textbf{(a)} Target present \hspace{7cm} \textbf{(b)} Target absent
\end{flushleft}
\vspace{-0.7cm}
    \begin{center}
      \includegraphics[width=0.4\linewidth]{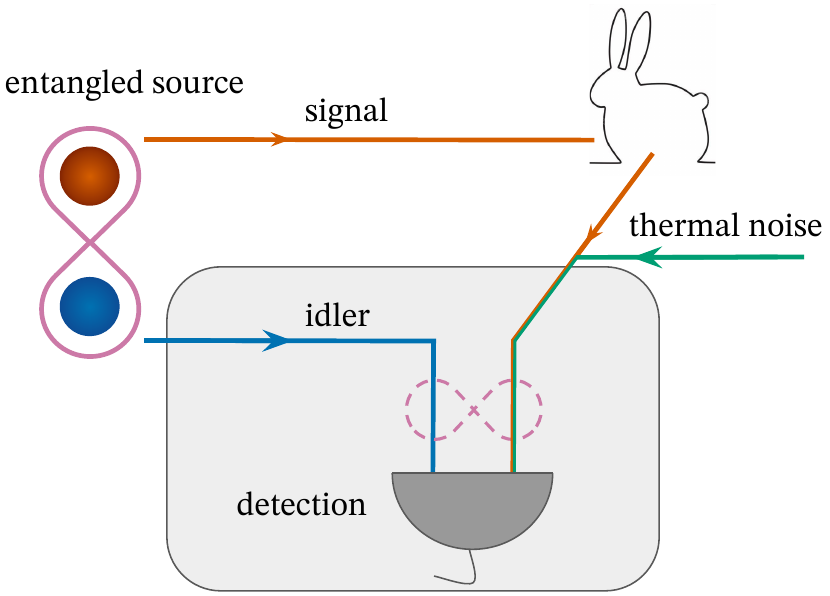}  
    \hspace{2cm}
      \includegraphics[width=0.4\linewidth]{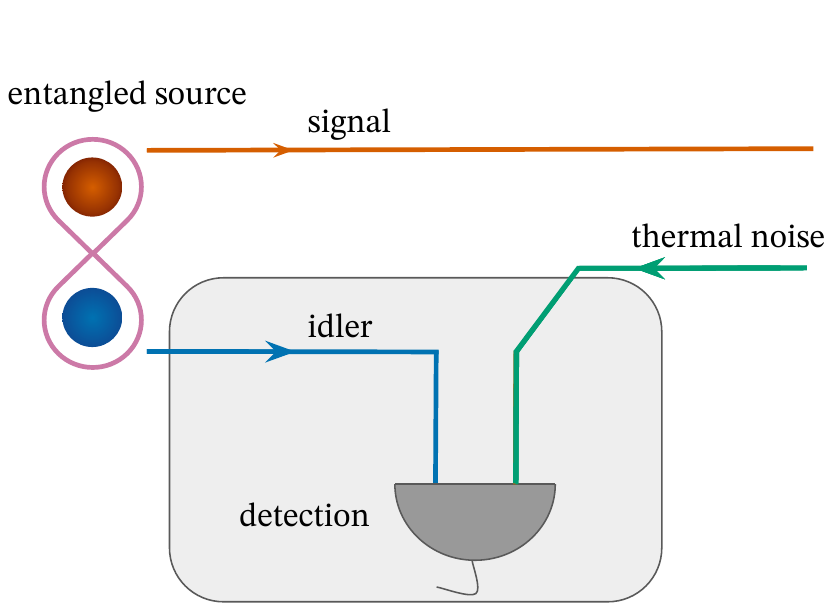}  
    \end{center}
    \caption{\textbf{Basic paradigm for quantum illumination.} A source produces an entangled state, comprised of signal and idler. The idler is retained \cdel{and}\cadd{at} the detector and the signal is sent towards a possible target. If the target is present as in panel (a), an optimal joint measurement is tasked with detecting the signal and the idler (together with additional thermal noise). If the target is not present or does not reflect the signal as in panel (b), the detection is of the idler and background thermal noise only.}
    \label{fig:basic_paradigm}
\end{figure*}

\section{Basic Paradigm for Quantum Illumination}
\label{sec:basic}
\ifx
\textcolor{red}{Basic generic notions of a quantum illumination protocol. Nice, easy to understand diagram highlighting signal, idler and measurement. I've added this section to explain the basic idea on a high level. I think this would make the following discussion in criteria for quantum radar easier to follow and gives a nice gentle introduction to the topic. We can remove if you both think this is superfluous - AF. }\textcolor{purple}{No, not superfluous. I think this is a good idea -AK}
\fi

We begin with a high-level overview of the basic operation of common QI protocols. The basic paradigm is illustrated in Fig.~\ref{fig:basic_paradigm}. A source generates entangled quantum states consisting of, in the case of discrete systems, a pair of entangled qubits(dits); in the continuous variable case, two entangled modes. One part of the state, the \textit{signal}, is sent towards a region of space in which a target may or may not reside and the remaining part, the \textit{idler}, is retained at the detector. Ideally, if the target is present in that space the signal is reflected back towards the detector to be combined with the waiting, retained idler. If the target is not present, the signal is lost and only the idler remains. In either case, thermal background noise is introduced from the environment and combined with the modes at the detector. An optimal joint measurement \cadd{(one which simultaneously measures both the signal and the idler) }
is performed on the states at the receiver and the measurement statistics are used to determine the possible presence of the target. The problem becomes equivalent to one of hypothesis testing, in which we wish to use the measurement data to discriminate between the two scenarios: target present or target absent.

\section{Quantum Radar Types and Criteria}
\label{sec:criteria}
It is often implicitly assumed that quantum radar is simply a \cdel{quantum illumination (QI)}\cadd{QI}
 scheme applied to the problem of distant target detection. Indeed, we have ourselves been guilty of being rather loose on the distinction thus far in this review. Let us now pause briefly to be a little more precise and consider the differences between quantum radar and QI and give a general definition of the former.

It is certainly still the case that a radar scheme relying on an entangled source and idler, that is a scheme based on QI, can safely be described as quantum radar. However, a more general definition of these technologies ought to be viewed as: \textit{any target detection scheme that employs any non-classical part for the purposes of enhanced capabilities}. The part which renders the device the title of being `quantum' may be in the form of a non-classical transmitter, a non-classical receiver, or both. Succinctly, one\cadd{'}s choice of quantum resource yields a quantum sensor that falls into one of three main types\cadd{~\cite{harriscorp2009}}:

\begin{itemize}
    \item \textbf{Type 1:} Non-classical quantum states of light are transmitted which are not entangled with the receiver. E.g., single-photon transmitters.
    \item \textbf{Type 2:} Classical (coherent) states are transmitted but quantum receivers are used to increase sensitivities. 
    \item \textbf{Type 3:} Quantum states of light are transmitted which are initially entangled with the receiver. (Joint quantum measurements are correspondingly used.)
\end{itemize}

At this point it is clear that the scope of quantum radar spans far beyond QI which itself may be classified as a Type 3 quantum sensor, as it entails generating entanglement between two modes: one is employed as the signal while the other, the idler, forms part of the receiver. Nonetheless, QI has served as an instigator in the creation of wide interest in quantum enhanced sensing schemes and is therefore largely responsible for the advances made to date in our understanding of quantum radar more generally. 

Regardless of the type of protocol utilised, several criteria must be met for a practical quantum radar to operate with quantum advantage. \cadd{Furthermore, since quantum radar is a continually}\cdel{As an} evolving and emerging technology\cadd{,} it \cdel{is sometimes}\cadd{can be} difficult to be sure whether a given result would qualify\cadd{ as a} proof of principle demonstration\cdel{ of a quantum radar}. The criteria below set out the fundamental requirements for a quantum target detection protocol to constitute a quantum radar.
\begin{itemize}
    \item \textbf{Outperform classical benchmark.} Clearly, the quantum protocol must outperform the equivalent optimal classical protocol in some realistic domain. (These classical benchmarks are discussed in Sec.~\ref{sec: theory}.)
    \item \textbf{Operates in ambient environment.} A realistic quantum radar should show quantum advantage in natural environmental conditions of temperature (in particular, beyond cryogenic temperatures for the signal systems).
    \item \textbf{Ranging capabilities.} Besides improved target detection, the quantum protocol should also be able to determine the distance of the target.
    \item \textbf{Radar-like distances.} Typically radars are used for target detection over relatively long distances.
    \item \textbf{Fast detection and data processing.} A quantum radar should provide an answer on a time scale which is much faster than the typical speed of the target.
\end{itemize}
To date, there has been no experimental implementation which completely satisfies these criteria. The criterion which has been (partially) demonstrated is some form of quantum advantage over corresponding classical benchmarks. Such advantage was investigated in short-range room temperature scenarios (of the order of 1 meter)~\cite{chang2019quantum,luong2019receiver,shabirQI}, or in artificial cryogenic conditions~\cite{Huard2022}. All cases considered fixed targets with no ranging capabilities and long detection times.

\section{Theoretical notions}
\label{sec: theory}

\subsection{Classical detection theory}
\label{subsec: classical}

Before proceeding with the discussion of quantum radar, it is important to understand some basic concepts about the current technology some feel it is destined to replace. Radar is a sensing technology originally developed during the first half of the 20th century, mostly just after the end of the First World War, even though reflection of radio waves by solid objects was first observed by Heinrich Hertz in 1886~\cite{richards2005fundamentals}. The first pulse-based radar system was developed by the US Naval Research Laboratory in 1934, able to detect and estimate the range of a target. Radar has become a major field of research and development, but its basic underlying principles remain mostly unchanged since its inception and the beginning of its widespread use more than 50 years ago. 

\subsubsection{The radar equation}
Let us consider the generic operation of a classical monostatic radar system, with co-locating transmitter/receiver. The transmitter emits an electromagnetic pulse of power, $P_{\mathrm{tx}}$, towards a target located some distance $R_{\mathrm{T}}$ away. The antenna does not emit radiation isotropically; it is instead usually emitted as a narrow beam with an additional element of directionality characterised by the transmitter gain, $G_{\mathrm{tx}}$. The gain is generally a function of spherical coordinates $\theta$ and $\phi$ and is a product of two terms that determine actual transmitted power: 1) the efficiency of the radar in generating a transmission signal from its input, and 2) the directivity of the antenna in terms of the actual outgoing beam. Directivity is calculated such that a perfectly isotropic antenna will have directivity of 1 in all directions and, even for those that are not isotropic (as in the case of pencil and fan beams), the mean directivity is still always 1 but varies with direction, having minimum and maximum values that are at most or at least 1, respectively. This multiplicative factor constituting the antenna gain is used to obtain the total transmitted power in the target's direction, $P_{\mathrm{tx}}G_{\mathrm{tx}}$ (assuming the peak gain is directed towards the target). The emitted electromagnetic pulse, after reflection off the target, arrives back at the receiver after a time, $\Delta t$, which can be used to compute the target range, $R = c \Delta t/2$, where $c$ is the speed of light in a vacuum.

In general, the majority of the emitted energy is lost during this process. This is due to attenuation factors associated with the medium through which the pulse propagates, and also the reflectivity of the target itself, which is dependent on both its material structure and its geometry. As a result, the total power density incident on the target is given by
\begin{equation}
W_{\mathrm{T}} =\frac{ P_{\mathrm{tx}} G_{\mathrm{tx}}  F^2}{4\pi R_{\mathrm{T}}^2},
\end{equation}
where $F \in [0,1] $ is a form factor describing the transmissivity of the space between the radar and the target, and the additional factor of $1/4 \pi R_{\mathrm{T}}^2$ describes the loss due to the pulse propagating as a spherical wave.

The reflectivity of the target, described by a single term called the radar cross-section, $\sigma$, quantifies the proportion of incident power that is subsequently scattered. This then propagates back towards the receiver such that the total power density arriving back at the receiver is given by
\begin{equation}
W_{\mathrm{r}} = \frac{ P_{\mathrm{tx}} G_{\mathrm{tx}}  \sigma F^4}{(4\pi)^2 R_{\mathrm{T}}^4} = \frac{P_{\mathrm{r}}} {a_{\mathrm{r}}},
\end{equation}
where $P_{\mathrm{r}}$ is the power arriving at the receiver and $a_{\mathrm{r}}$ is the receiver's collecting area. Thus we arrive at the radar equation 
\begin{equation}
P_{\mathrm{r}} = \frac{P_{\mathrm{tx}} G_{\mathrm{tx}} a_{\mathrm{r}} \sigma F^4}{(4 \pi )^2 R_{\mathrm{T}}^4},
\label{eq:RadarEquation}
\end{equation}
which is the fundamental model for classical radar systems.
Note that the use of a single value for radar cross section is a simplification. This cross section varies with the frequency of the radar wave, the target's orientation, and its specific configuration. These factors affect how radar waves interact with the target, leading to significant variability in radar return signals.

\cadd{Finally, we note that any future quantum radar would still be subject to the radar equation, and the received signal strength would have the same $1/R_{\mathrm{T}}^4$ dependence to the target's range. As will be discussed in later sections, QI-based quantum radars would therefore require an \emph{extremely} high power 
in order to be useful at large ranges. Based on current technological capabilities, achieving such a feat would be prohibitively costly and/or wildly impractical. Nevertheless, since microwave quantum technologies is a broad-scoped field which is continually making fast-paced progress, achieving such a goal is not totally out of the question in the future.}

\subsubsection{The signal-to-noise ratio (SNR)}

Any radar receiver is inherently susceptible to thermal noise which places limitations on the strength of target signals that can be detected \cite{
skolnik2008introduction,jenn2005radar}. \cadd{Johnson–Nyquist noise arises from electronic noise intrinsic to the radar system (thermal vibrations of charge carriers, independent of the applied voltage).} Its total associated power is given by
\begin{equation}
P_{\mathrm{n}} = k_B T B_{\mathrm{n}} F_{\mathrm{n}},
\label{eq:thermalnoise}
\end{equation}
where $k_B$ is Boltzmann's constant, $T$ is the system's operating temperature, $B_{\mathrm{n}}$ is the receiver's bandwidth and $F_{\mathrm{n}}$ is a dimensionless constant expressing the variation of the true noise characteristics from that of an ideal black body. Using Eq.~(\ref{eq:thermalnoise}) we may \cdel{can} then define the signal-to-noise ratio (SNR) as
\begin{equation}
\mathrm{SNR} = \frac{P_{\mathrm{r}}}{P_{\mathrm{n}}} =  \frac{P_{\mathrm{tx}} G_{\mathrm{tx}} a_{\mathrm{r}} \sigma F^4}{(4 \pi )^2 R_{\mathrm{T}}^4 k_B T B_{\mathrm{n}} F_{\mathrm{n}}}.
\label{eq:SNReq1}
\end{equation}
The \cadd{(effective)} collecting area of the receiver, $a_{\mathrm{r}}$, can be further expressed in terms of the receiver gain, $G_{\mathrm{r}} = 4\pi a_{\mathrm{r}} / \lambda^2$, with $\lambda$ being the wavelength. This allows us to rewrite Eq.~(\ref{eq:SNReq1}) as
\begin{equation}
\mathrm{SNR}  =  \frac{P_{\mathrm{tx}} G_{\mathrm{tx}} G_{\mathrm{r}} \lambda^2 \sigma F^4}{(4 \pi )^3 R_{\mathrm{T}}^4 k_B T B_{\mathrm{n}} F_{\mathrm{n}}}.
\label{eq:SNReq2}
\end{equation}
Since the SNR always has a finite value, from Eq.~(\ref{eq:SNReq2}) there exists a minimum detectable signal, $\mathrm{SNR}_{\mathrm{min}}$, which must exceed the system's noise floor. This in turn corresponds to a maximum detection range,
\begin{equation}
R_{\mathrm{max}} \simeq \left(  \frac{P_{\mathrm{tx}} G_{\mathrm{tx}} G_{\mathrm{r}} \lambda^2 \sigma F^4}{(4 \pi )^3 k_B T B_{\mathrm{n}} F_{\mathrm{n}} \mathrm{SNR}_{\mathrm{min}}}   \right)^{1/4},
\end{equation}
where $\mathrm{SNR}_{\mathrm{min}}$ is typically of the order of 10-20 dB \cite{jenn2005radar}.



\subsubsection{Hypothesis testing}\label{sec:classicalhypothesistesting}
Radar detection entails, with the view of best describing a detected signal, making a decision between two possible hypotheses~\cite{lehmann2006testing}: The null hypothesis $H_0$, corresponding to target absence, and the alternative hypothesis $H_1$, corresponding to target presence~\cite{richards2005fundamentals}. This simple example of a binary decision task has been the subject of many studies and its analysis begins with the definition of two probability density functions (pdfs) describing the measurement to be tested under each of the two available hypotheses. Supposing the sample to be tested is denoted $x$, then we need two pdfs:
\begin{equation}
\begin{split}
p_x(x|H_0) &= \text{pdf of $x$ given that the target was absent}, \\
p_x(x|H_1)& = \text{pdf of $x$ given that the target was present}.
\end{split}
\end{equation}
These may be generalised to the $M$-dimensional joint pdfs $p_{\bm{x}}(\bm{x}|H_0)$ and $p_{\bm{x}}(\bm{x}|H_1)$ for a detection problem based on $M$ i.i.d samples of data $x_n$ forming the sample vector $\bm{x} \equiv \left[ x_1 \dots x_M \right] ^T$, with $n = 1, \dots, M$. The underlying problem is reliant on the proper modelling of these two functions and their estimates are in turn reliant on the system and parameters governing the scenario in question. 

Based on the above pdfs, the following probabilities of interest may be defined:
\begin{itemize}
\item Probability of detection, $P_{\mathrm{det}}$, \\The probability of a target being correctly declared present, $P(H_1|H_1)$;
\item Probability of false alarm (Type I error), $P_{\mathrm{fa}}$, \\The probability of a target being incorrectly declared present, $P(H_1|H_0)$;
\item Probability of mis-detection (Type II error), $P_{\mathrm{md}}$, \\The probability of a target being incorrectly declared absent, $P(H_0|H_1)$.
\end{itemize}
Optimisation of these probabilities can be carried out in a range of ways based on the rules one wishes to follow for decision making, which may also be situation dependent. \cadd{This is a very rich field, which can may explored in great detail through Refs.~\cite{johnson1993array} and~\cite{kay1998fundamentals}.} In radar, a common choice is the Neyman-Pearson criterion in which the probability of detection, $P_{\mathrm{det}}$, is maximised under the constraint that the Type II false alarm probability does not exceed some predetermined, tolerable value. The overall diagnostic capability of binary classifiers, such as radars, can be evaluated by the construction of a receiver operating characteristic (ROC) curve\cadd{~\cite{richards2005fundamentals}}. This graphically plots the true detection rate $P_{\mathrm{det}}$ against the false alarm rate which is generally fixed as part of the system specifications\cdel{~\cite{richards2005fundamentals}}, and shows the trade-off between sensitivity ($P_{\mathrm{det}}$) and specificity (1-$P_{\mathrm{fa}}$). A commonly used measure of performance is given by the area under the curve, which may be interpreted as the probability that the model ranks a random positive example more highly than a random negative example.

Assuming that the probabilities, $p_0$ and $p_1$, associated with the two hypotheses are known (the so-called `prior probabilities' or `priors'), one can consider a symmetric version of the problem, where the two hypotheses have the same Bayesian cost. In this case, the mean error probability associated with the target detection may be written as $p_{\text{err}}= p_0 P(H_1|H_0) + p_1 P(H_0|H_1)$. While the scenario of known priors is not typically considered in classical detection theory, it represented a simplified framework for testing the performance of the new quantum protocols. The symmetric formulation allowed scientists to show quantum advantage by analysing a single figure of merit.

\subsection{Quantum illumination}
\label{sec:QI}
\subsubsection{Discrete-variable QI}
QI, in its original description, refers to the use of discrete-variable non-classical states to enhance the detection of potentially remote objects. In 2008, Ref.~\cite{lloyd2008enhanced} demonstrated that the use of entangled pairs of qubits could enhance the detection probability of a low-reflectivity target. Two protocols were considered and compared: in the first, the beam was composed on $N$ unentangled single-photon states and, in the second, two entangled beams (labelled signal and idler) were generated. In both scenarios, the optical transmitter illuminated a region of space in which a weakly-reflecting target was equally likely to be present or absent. In both the entangled and unentangled cases, the assumptions made were as follows:
\begin{itemize}
    \item Signal comprising $N$ high time-bandwidth product $M=TW \gg1$, single-photon pulses. Here, $T$ is the detection time window and $W$ is the bandwidth such that the detector can distinguish between $M$ modes per detection event.
    \item Round-trip transmissivity $0 < \kappa \ll 1$ when the target is present; $\kappa=0$ when the target is absent.
    \item Low background noise $N_B \ll 1$.
    \item For each transmitted signal pulse, at most one photon is detected at the receiver such that $M N_B \ll 1$.
\end{itemize}
Under these assumptions, two regimes, ``good'' and ``bad'', were identified for the operation of each of the single-photon (SP) and entangled QI sources based on the values computed via the quantum Chernoff bound (QCB)~\cite{Audenaert2007QChernoff} (which is an asymptotically tight upper bound for the mean error probability in the symmetric formulation of quantum hypothesis testing). In their good regimes, the average probabilities of making an error after $N$ trials were estimated to be\cadd{~\cite{shapiro2009quantum}}
\begin{equation}
    P_{\mathrm{err}}^{\mathrm{SP}}\simeq\frac{1}{2}e^{-\kappa N},\, \kappa \gg N_B,
\end{equation}
and
\begin{equation}
    P_{\mathrm{err}}^{\mathrm{QI}}\simeq\frac{1}{2}e^{-\kappa N},\, \kappa \gg \frac{N_B}{M},
\end{equation}
respectively, showing that the QI case afforded a much larger region of validity. In their bad regimes, the estimates for the error probabilities via the QCBs are instead given by\cadd{~\cite{shapiro2009quantum}} 
\begin{equation}
    P_{\mathrm{err}}^{\mathrm{SP}}\simeq\frac{1}{2}e^{-\frac{\kappa^2 N}{8 N_B}},\, \kappa \gg N_B,
\end{equation}
and
\begin{equation}
    P_{\mathrm{err}}^{\mathrm{QI}}\simeq\frac{1}{2}e^{-\frac{\kappa^2 N M}{8 N_B}},\, \kappa \gg \frac{N_B}{M},
\end{equation}
yielding, again, an enhancement in the region of validity for QI with the error probability drastically reducing for $M \gg 1$ in comparison to unentangled single-photon sources.

These results rely on various, typically unrealistic, assumptions. Firstly, one requires the existence of a source of entangled high time-bandwidth product photons to probe the target region. Upon their eventual return, one demands that the receiver is an optimal one, performing an optimal joint measurement on all of the returning photon with their corresponding idlers, all the while ensuring that the idler storage system was completely lossless throughout the signal's round-trip. These limitations have been the subject of much debate regarding how realistic a practical QI-based quantum radar would be and will be explored in further detail later in this review.

Of course, it is important to note that this initial comparison was not entirely comparing a quantum scheme to a classical one. In fact, it compared one using entanglement to one that did not. The use of single photons as a source for target detection, while not being entangled at all, still requires the use of quantum photo-detection theory in order to process the data and formulate a decision. Indeed, this is simply a form of Type 1 quantum sensor described in Sec.~\ref{sec:criteria}.

Instead a comparison should be made to a classical equivalent as a benchmark. To clarify, this does not refer to using the classical radar described in Sec.~\ref{subsec: classical} as a benchmark, but instead to a comparison between the performance of the QI protocol and the same protocol performed with the most ideal, `classical state' within quantum optics: the coherent state.
Further, since coherent states are Gaussian, their use in quantum optics experiments can be straightforwardly studied and modelled in frameworks which mirror those making use of true quantum mechanical phenomena. This provides a means for formally defining and isolating a quantum advantage. The performance of a QI protocol using such coherent states is frequently referred to as the ``classical benchmark'', at least in the optical domain.

In 2009, Ref.~\cite{shapiro2009quantum} provided such a comparison between entanglement-based QI and the equivalent coherent state protocol. Here, the single-photon transmitter was replaced with a coherent state transmitter with $N$ pulses, each with average photon number of unity. In this case, the QCB yields\cadd{~\cite{shapiro2009quantum}}
\begin{equation}
    P_{\mathrm{err}}^{\mathrm{CS}}\leq \frac{1}{2}e^{-  \kappa N (\sqrt{N_B + 1}- \sqrt{N_B})^2},
\end{equation}
applicable for all values of $0 \leq \kappa \leq 1$ and $N_B \geq 0$. In the regime of low-background, i.e., $N_B \ll 1$, this reduces to\cadd{~\cite{shapiro2009quantum}}
\begin{equation}
    P_{\mathrm{err}}^{\mathrm{CS}}\leq \frac{1}{2}e^{-  \kappa N}.
\end{equation}
The performance of the coherent state protocol therefore matches Lloyd's previous ``good'' regime QI performance while substantially outperforming discrete-variable QI in its ``bad'' regime. Thus, although entangled photons outperformed the single photon equivalent, no quantum advantage had yet been found over the classical benchmark.

\subsubsection{Gaussian QI}
\label{sec:GaussQI}

With the results of Ref.~\cite{shapiro2009quantum}, for a brief period, QIs prospects fell flat with its potential appearing limited at best. Fortunately though, at around the same time of Lloyd's work, Tan \etal~\cite{tan2008quantum} published the continuous-variable version of QI involving the use of Gaussian states. (We refer interested readers to Refs.~\cite{braunstein2005quantum,weedbrook2012gaussian,serafiniBook} for comprehensive reviews on Gaussian quantum information and continuous-variable systems.)
 
Consider a resource state modelled as a two-mode squeezed vacuum (TMSV) Gaussian state comprising a signal mode sent out to some target region and an idler mode retained at the source for later joint measurement, each with $N_S$ photons per mode. The theory of Gaussian QI assumes the following conditions:
\begin{itemize}
     \item Low-brightness signal, $N_S \ll 1$.
     \item High time-bandwidth product, $M=TW \gg 1$.
     \item Low target reflectivity, $0\leq \kappa \ll 1$ (with $\kappa =0$ when the target is absent).
     \item High-brightness thermal background, $N_B \gg 1$.
\end{itemize}

Two alternate hypotheses exist for the experiment's outcome. The first, $H_0$, with the target absent, where the returning signal is just a noisy background modelled as a thermal state with mean number of photons per mode $N_B \gg 1$ (note the difference here to previous works where the assumption was that $N_B \ll 1$ and $M N_B \ll 1$). The second case, $H_1$, corresponds to a weakly reflective target being present in the region with reflectivity, $\kappa \ll 1$, giving the proportion of signal modes reflected back towards the source, physically representing a high loss regime. This is combined with a very strong background, now with mean photons per mode of the return given by $N_B / (1-\kappa)$. In either case, the returning signal and the retained idler are no longer entangled. 

The decision problem is reduced to one of being able to distinguish between the two conditional states at the receiver and our ability to do this can be quantified via the computation of various bounds. Our choice of such bound depends on how we wish to weight the associated costs for error types and are reduced to the consideration of symmetric or asymmetric costing procedures.
Ref.~\cite{tan2008quantum} considered the setting of symmetric quantum hypothesis testing (QHT) and found that the mean error probability is asymptotically bounded as follows
\begin{equation}\label{QIbound}
P_{\mathrm{err}}^{\mathrm{QI}} \leq  \frac{1}{2} e^{- M \kappa N_S/N_B},
\end{equation}
in the limits $0<\kappa \ll 1$, $N_S \ll1$ and $N_B \gg 1$.

Meanwhile, Ref.~\cite{tan2008quantum} found that the corresponding $M$-pulse coherent-state source (yielding an optimal classical benchmark) has a QCB given by
\begin{equation}\label{CSboundexact}
P_{\mathrm{err}}^{\mathrm{CS}} \leq \frac{1}{2} e^{- M \kappa N_S \left(\sqrt{1+N_B} - \sqrt{N_B} \right)^2},
\end{equation}
valid for all parameter values. Imposing the same limitations used in the derivation of Eq.~(\ref{QIbound}), that is, for $0<\kappa \ll 1$, $N_S \ll1$ and $N_B \gg 1$, the coherent-state transmitter QCB, in such a regime, is given by
\begin{equation}\label{CSbound}
P_{\mathrm{err}}^{\mathrm{CS}} \leq \frac{1}{2} e^{- M \kappa N_S/4 N_B}.
\end{equation}
In contrast to earlier results, the QI transmitter's error exponent in this regime has a factor of 4 (equivalent to 6 dB) advantage over the corresponding coherent-state transmitter.
 
 It has been shown that this 6 dB advantage is theoretically maximal~\cite{di2021two} under the consideration of optimal collective quantum measurements across all $M$ of the employed mode pairs. This is reduced to 3 dB in the presence of more practical receiver designs, for instance if the receiver is restricted to local operations and classical communications and joint measurements are carried out only on individual mode pairs (see also Ref.~\cite{karsa2020noisy}). \cadd{The fundamental limits of QI has also been shown in Ref.~\cite{nair2020fundamental} where for $N_B \gg 1$ and $N_S \ll 1$ the TMSV is shown to achieve the greatest quantum mechanically allowable error probability exponent. They further show that near optimality in detection persists for targets exhibiting flat Rayleigh fading, i.e., when the reflectivity is exponentially distributed about the targets average reflectivity and the phase of the returning signal beam is uniformly distributed over $\left[0,2\pi\right)$. This optimality was also shown in Ref. ~\cite{bradshaw2021optimal} for the detection of targets with vanishing reflectivity: for single-mode Gaussian probes, the optimal state is the coherent state; for two-mode Gaussian probes, the optimal state is the TMSV.}
 
 Further, the two-mode Gaussian state considered in achieving these bounds has been shown to be the optimal quantum state~\cite{de2018minimum} in the context of asymmetric QHT (meaning that no other, more exotic, quantum states such as ones exhibiting multi-mode entanglement can improve target detection). The same work also showed that, in the absence of a quantum memory, i.e., with no ability to store an idler, the coherent state forms the optimal single-mode source. The latter is true when the energy constraint is local, i.e., we fix the mean number of photons per signal. When the energy constraint is global (i.e., we fix the total number of photons carried by all the signals to some fixed amount), then the use of coherent states can be outperformed by squeezed states whose displacement and squeezing are jointly optimised~\cite{spedalieri2021optimal}.

\subsubsection{Microwave QI}\label{sec:2_2microwaveQI}
The main findings of Refs.~\cite{lloyd2008enhanced,tan2008quantum} presumed operation at optical wavelengths. In the optical domain, the necessary tools for QI implementation are well-known and widespread, including the use of spontaneous parametric down-conversion (SPDC) sources for entanglement generation (which naturally produce low-energy modes $N_S \ll 1$) and high fidelity single-photon detectors which are largely quantum-limited in noise associated with their operation. While this assumption poses no issue for the setup of the QI protocol, the result of a 6 dB quantum advantage by Tan \etal\ was found under the assumption that the ambient background mean number of photons per mode is $N_B \gg 1$. This condition does not occur naturally at optical wavelengths where, in fact, $N_B \sim 10^{-6}$ and smaller. In fact, it has been shown that in the absence of noise the optimal quantum radar's performance is closely approximated by a conventional coherent state radar operating at the same transmitted energy~\cite{nair2011discriminating}.

A natural solution to this issue would be to extend the theory of QI to the microwave domain where the thermal background provides the necessary $N_B \gg 1$ for quantum advantage. Ref.~\cite{barzanjeh2015microwave} successfully achieved this extension \cdel{starting from a standard SPDC photon source to generate entangled signal-idler optical mode pairs. The protocol proceeded with the employment of an electro-optomechanical (EOM) converter to down-convert the optical signal into a microwave mode which was then transmitted to the target region of interest.}\cadd{by employing electro-optomechanical (EOM) converters to mediate interactions across different frequencies of light. A first EOM module directly generates optical-microwave entanglement by exploiting the interaction mediated by the mechanical oscillator to generate the propagating microwave signal.} Upon \cdel{the signal's}\cadd{its} return, a further EOM converter \cdel{wa}\cadd{i}s used to reverse the process, up-converting back to the optical region to undergo a phase-conjugated (PC) joint measurement with the retained optical idler. It is this microwave extension to QI which led to it becoming of great interest in the wider community as a potential quantum-mechanical alternative to the classical radar, potentially enabling one to detect a \cdel{stealth}\cadd{low-visibility} target, while hiding a weak signal in a naturally occurring strong background.

\subsection{Benchmarking quantum target detection protocols}\label{sec:benchmarking}

\cadd{Recent years have seen demonstrations the first prototype microwave QI experiments~\cite{luong2019receiver,shabirQI}. Reported quantum enhancements in SNR was relative to their chosen classical comparison case, benchmarked within their specific experimental set-up, which was not the same as the `optimal' classical benchmark based on coherent states which is widely assumed. As a result, these reports have been the subject of much debate across the wider research community since, in essence, their chosen classical benchmarks were sub-optimal. However, it is important to point out that within the setting of a room-temperature microwave illumination experiment, there are very few known methods to effectively generate a low-energy coherent microwave source. As has been previously outlined in Ref.~\cite{karsa2021classical}, there are three potential procedures for benchmarking quantum radar protocols. The possibilities involve generating the semi-classical source with (1) an amplifier (realistic); (2) cryogenic attenuation (realistic); and (3) neither of these additional elements (ideal). }

\cadd{In any case, the classical benchmark should take the form of a room-temperature microwave coherent state with a low mean number of photons per mode ($N_S \ll 1$) to compare with the TMSV state of the same energy. Amongst the three possible generation methods outlined, only the latter one is capable of generating a perfect coherent state, but this requires availability of a device which reliably generates low-energy coherent states at microwave wavelengths. This would yield the theoretically optimal classical source and, in turn, the ideal classical benchmark to model microwave QI experiments against. Unfortunately, such a device does not exist; one has to instead resort to one of the first two possibilities, or some hybrid scheme mixing the two.}

\cadd{Possibility (1) entails the generation of a low-energy microwave coherent state, which is possible at ultra-low temperatures ($\sim 7$mK). The resulting signal must be passed through an amplifier in order to probe a target region at room temperature ($\sim 300$K), also compensating for detector limitations and free-space loss. The necessary use of an amplifier in this option unavoidably introduces noise to the state, such that the resultant classical probe is not a coherent state and therefore sub-optimal. Possibility (2) is based on recent developments in the production of room-temperature `microwave lasers' (masers) in solid-state devices~\cite{breeze2018continuous,wu2020room}. While these readily generate microwave coherent states, their energies are much too high ($N_S \gg 1$) to allow for sensible benchmarking of QI. This may be remedied by using heavy cryogenic attenuation to reduce the signal energy while minimising the added noise which, for sufficiently low-temperature attenuation, can maintain an approximate coherent source. Such a scheme has not, as of yet, been experimentally demonstrated. Note that Ref.~\cite{shabirQI} used a method which was a hybrid between (1) and (2): a room-temperature microwave source generated a weak coherent tone which was then subject to a series of low-temperature attenuators. Amplification followed to enable returning signal detection.}

\cdel{The first microwave QI experiments have been carried out showing improved performances over their chosen classical comparison cases. This has been the subject of much debate, since these classical comparison cases are indeed different to the traditionally `optimal' one based on coherent states and, as such, their performances may be viewed as sub-optimal. However, there are very few known methods for generating a low-energy semi-classical microwave source for room temperature applications. Currently, there are three potential procedures, as pointed out in Ref.~\cite{karsa2021classical}:}

\cdel{Note that the source generation method used in the prototypical experiment~\cite{shabirQI} was in fact a hybrid between procedures (1) and (2): a room temperature microwave source generated a weak coherent tone followed by a chain of low temperature attenuators which was then amplified to enable returning signal detection. Further, d}Despite the fact that \cadd{with current capabilities} \cdel{procedure}(3) is impossible to carry out\cdel{with current experimental capabilities}, it persists to be assumed as the single classical benchmark in almost all literature pertaining to microwave QI. While it is certainly valid and optimal in the optical regime, this does not \cdel{translate to}\cadd{hold true in} the microwave where it simply does not exist. Knowledge of the true, regime-dependent, classical benchmark is crucial in order to ascertain the existence of \cadd{and quantify} a quantum advantage. The community should start to be more precise in terms of classical benchmarks and start to distinguish between `relative' quantum advantage, with respect to the realistic benchmarks at points (1) and (2) above, and `absolute' quantum advantage, with respect to the ideal benchmark at point (3).

\ifx
This chapter outlines a true classical benchmark for microwave QI for room temperature applications, based on the fact that these techniques are, so far, the only known tools for generating an optimal classical source at the microwave. Sec.~\ref{sec:protocol} outlines two protocols for microwave QI using coherent states: the first, for a source generated with amplification; the second, proposed by this work, based on the output of a room temperature maser followed by heavy cryogenic attenuation. The tools of quantum hypothesis testing (QHT) are used in Secs.~\ref{sec:symmetric} and~\ref{sec:asymmetric} where formulae for the quantum Chernoff bound (QCB) and quantum relative entropy (QRE) are given, under symmetric and asymmetric considerations, respectively, yielding new error bounds for the microwave classical benchmark. In Sec.~\ref{sec:homodyne}, a protocol involving homodyne detection of the returning signal is considered with the resulting receiver operating characteristic (ROC) computed. In all cases, the results for these new classical benchmarks are compared to the traditional one applicable only in the optical regime, constrained such that the total energy by which the target is irradiated is maintained. Up to here, this work's analyses are confined to regimes whereby the simultaneous study and comparison of classical benchmarks (1), (2) and (3) are possible. The sheer magnitude of the noise introduced by procedure (1) renders the signal energy per mode so large that any quantum advantage would be diminished owing to the fact that the two-mode, signal-idler, entanglement correlations enabling the QI advantage become irrelevant (see Sec.~\ref{sec:32gensource}) at high brightness. Thus, in Sec.~\ref{sec:analysis} the results of Sec.~\ref{sec:classbench} are studied as the classical benchmark and compared to the performance of a two-mode squeezed vacuum (TMSV) source for entanglement-based QI, within the regime where such a protocol may be applied.
\fi

\section{Theoretical and Experimental Challenges}\label{sec:exp reqs}
We now turn to address the challenges for performing a Type 3 quantum radar protocol as defined in Sec \ref{sec:criteria}. We consider in turn, the generation of entangled modes at the source, the question of how to store the idler mode at the detector and design implications for the receiver. Additionally we briefly discuss the difficulties in extending from target detection to range finding in the quantum case. Note that some of these issues were first discussed in Ref.~\cite{pirandola2018advances}.

\subsection{Source generation at the microwave \cadd{wavelengths}}

QI relies on the generation of maximally entangled mode pairs with high time-bandwidth product. Microwave entanglement can be readily generated directly in superconducting circuits via Josephson parametric amplifiers~\cite{chang2018generating,yurke1989observation} which is exactly the method employed in prototypical experiments for microwave QI. However, the issue here is that these signals are very faint and their amplification for room temperature applications induces a critical degradation of their quantum features. It is worth noting that there exists \emph{some} flexibility in the quality of quantum correlations required in order for a quantum advantage to be possible, as shown in Ref.~\cite{karsa2020generic}. \cadd{This work analysed the potential of loosening the transmitter requirements of QI, to enable one to consider a Gaussian source whose quantum correlations could vary in value, resulting in two-mode quantum states varying between just-separable to maximally entangled. The results were dependent on the form of hypothesis testing considered; when asymmetric, allowing some tolerable value of $P_{\mathrm{fa}}$ while further minimising $P_{\mathrm{md}}$, a range of non-maximal values of quantum correlations was shown to achieve a quantum advantage under realistic applications.} Though, of course, this must be balanced with other sources of imperfection and inefficiency, such as idler storage and receiver design.

In the optical domain, however, SPDC sources can reliably output the required photon numbers needed in order to perform QI. While the envisioned EOM converter of Ref.~\cite{barzanjeh2015microwave} \cadd{(see also Sec.~\ref{sec:2_2microwaveQI})} could provide a solution here by means of frequency conversion, such technologies are yet to reach the required specifications for experimental implementation in the context of quantum radar. The associated direct-conversion transduction device, ultimately reducing to a beamsplitter-type interaction between different frequency states, has been subject to long term active development owing to its vital role for scaling quantum computers and quantum networks. While great steps have been taken via a host of potential platforms~\cite{han2021microwave}, reported direct-conversion efficiencies consistently fall short of the $\eta_{\mathrm{DC}} \geq 1/2$ lower bound requirement for non-zero quantum capacity to yield entanglement transfer. Such an introduction of non-trivial resource degradation already at initial stages of transmission is crucial to overcome for QI applications.  

An alternative theoretical approach to electro-optical transduction has also been proposed based on continuous-variable quantum teleportation~\cite{wu2021deterministic}. It was shown that efficient microwave-optical transduction could be achieved along with a vast reduction of the strict thresholds required for direct-conversion systems to meet. Nevertheless, the scheme involves homodyne detection of the microwave modes which is, in itself, its own experimental challenge (discussed in Sec.~\ref{sec:benchmarking} regarding the ideal classical benchmark).

\cadd{The problem with microwave source generation, particularly creating microwave-microwave or microwave-optical entanglement is not limited to QI tasks. Quantum entanglement is a key ingredient for the success of many quantum technologies, and the ability to share this resource across different platforms is crucial to enable the development of optimised quantum networks which will inevitably require the assimilation of various platforms, potentially operating at different energy scales. Therefore, it is of great interest to the wider quantum technology community to bridge this energy gap (which may be of many orders of magnitude) to enable the realisation of future hybrid quantum systems. A recent demonstration~\cite{sahu2023entangling} used an optically-pulsed superconducting electro-optical device to create and verify entanglement between microwave and optical fields. Such a procedure can be used in place of the EOM devices of Ref.~\cite{barzanjeh2015microwave}, eliminating the mechanical part of the system, while still guaranteeing a scheme equivalent to the original.}

\subsection{Idler storage}\label{sec:idlerstorage}
Idler storage poses a critical problem for QI which must be overcome if one wishes to attain the maximum possible 6 dB enhancement. This performance gain hinges on the ability to perform an optimal joint measurement between the entirety of the returning signal and idler beams; losses during idler storage will hinder the amount of information attainable through the process and their recombination at a specific point in time is necessary for any successful joint measurement.

Suppose one's means of idler storage has an associated efficiency $\eta_{\mathrm{I}}$ quantifying the proportion of idlers successfully stored for later recombination with the returning signal. Then, the QCB for QI [Eq.~(\ref{QIbound})] would be modified as
\begin{equation}
    P_{\mathrm{err}}^{\mathrm{QI}} \leq \frac{1}{2} e^{-\frac{M \kappa \eta_{I} N_S }{N_B}}.
\end{equation}
Note that this means we require idler storage efficiencies $\eta_{I} \geq 1/4$ for any quantum advantage to be possible. In order to achieve any performances better than those obtainable via pairwise LOCC measurements (see also Sec.~\ref{subsec:QIreceivers}), requirement becomes $\eta_{I} \geq 1/2$. 

Possible means of idler storage are through optical fibre delay lines and quantum memories~\cite{heshami2016quantum}. The latter method, while potentially more efficient, is still a technology very much in its infancy.

\subsection{Designing a Gaussian QI receiver}\label{subsec:QIreceivers}
Having addressed the issues of generating a large number of quantum-correlated or entangled photon pairs and the storage of idlers during the experiment, it still remains to design a receiver capable of harnessing the QI advantage. After all, the quantum enhancement resides in the phase-sensitive cross-correlation terms existing between two different photonic modes\cdel{ and thus both must be measured simultaneously}. \cadd{This can be illustrated by considering the bosonic mode operators for the different modes and the relevant observables arising from them. For QI, this phase-sensitive term for the $k$th signal-idler mode pair is $\langle \hat{a}_R^{(k)} \hat{a}_I^{(k)} \rangle_{1} = \sqrt{\kappa N_S (N_S+1)} $ under $H_1$ (target present), while it is $\langle \hat{a}_R^{(k)} \hat{a}_I^{(k)} \rangle_{0} = 0$ under $H_0$ (target absent). This is where the target's signature resides, and its presence may be inferred by measuring the observable $\langle \hat{a}_R \hat{a}_I \rangle$.} \cadd{The required measurement is experimentally non-trivial. Any attempt to expand it in terms of quadrature operators, which may be measured individually, yields a result which means all must be known simultaneously.} Early attempts at designing a QI receiver focused on ones whose implementation entailed converting phase-sensitive correlations to phase-insensitive ones, which could then be subsequently subject to standard detection schemes. While able to promise some sort of quantum advantage, these fell short of the anticipated 6 dB advantage posited by Ref.~\cite{tan2008quantum}. Since then there has been a lot of research into Gaussian QI receiver designs\cadd{~\cite{guha2009gaussian,FFSFG,shi2023fulfilling,reichert2023quantum}}, becoming evermore promising in terms of performance and realisability. 

The following section will discuss most of the contributions to this area in turn, though, it is also important to note that these receivers would need to be implemented at the microwave frequencies (in case a frequency transducer is not used in the QI setup). It is also worth pointing out here that in realistic scenarios, \cadd{where there are imperfections in the reflecting surface (appearing rough relative to the scale of the transmitted wavelength) and variability in the pathway taken by each mode of light,} the amplitude and phase of the returning signal is randomly modified through \cadd{speckle and} fading~\cite{goodman1965,goodman1976}. This is particularly true for optical wavelengths. As a result of this, the quantum advantage afforded by some of the receivers outlined here can be severely degraded or completely nullified\cadd{, since they rely on detection of a signal with a known amplitude and phase~\cite{zhuang2017quantum}}.

\subsubsection{Optical parametric amplifying receiver}
Guha and Erkmen, in Ref.~\cite{guha2009gaussian}, first introduced a receiver based on the optical parametric amplifier (OPA), physically realisable by an SPDC crystal, which mixes input return and idler modes $\hat{a}^{(k)}_R$ and $\hat{a}^{(k)}_I$, $1 \leq k \leq M$, producing output mode pairs
\begin{equation}
\hat{c}^{(k)}=\sqrt{G} \hat{a}^{(k)}_I + \sqrt{G-1} \hat{a}^{\dagger (k)}_R,
\end{equation}
and
\begin{equation}
\hat{d}^{(k)}=\sqrt{G} \hat{a}^{(k)}_R + \sqrt{G-1} \hat{a}^{\dagger (k)}_I,
\end{equation}
where $G>1$ is the detector's gain which may itself be optimised to maximise QI reception. Then, assuming optimal photon counting on idler-output modes, they found the following upper bound for the mean error probability
\begin{equation}\label{QIOPAbound}
P_{\mathrm{err}}^{\mathrm{QI,OPA}} \leq \frac{1}{2} e^{- \frac{ M \kappa N_S}{2 N_B}},
\end{equation}
under the usual limits $0<\kappa \ll 1$, $N_S \ll1$ and $N_B \gg 1$. This yields \emph{at most} half of the ideal QI error exponent advantage, Eq.~(\ref{QIbound}), over coherent states, equivalent to 3 dB. 

\subsubsection{Phase-conjugating receiver}
Ref.~\cite{guha2009gaussian} also introduced the phase-conjugating (PC) receiver which, operating in the same regime as the OPA receiver, is capable of achieving the same 3 dB maximal enhancement in target detection. The returning modes are mixed with a vacuum mode $\hat{a}_V^{(k)}$ before conjugation to yield $\hat{a}_{PC}^{(k)}=\sqrt{2}\hat{a}_{V}^{(k)}+\hat{a}_{R}^{\dagger (k)}$. Then, recombination of the $k$th mode with its respective idler yields two output modes $\hat{a}_{\pm}^{(k)}=(\hat{a}_{PC}^{(k)} \pm \hat{a}_{I}^{(k)})/\sqrt{2}$. Detection proceeds with measurement of the operator $\hat{N} = \hat{N}_{+}-\hat{N}_{-}$, where $\hat{N}_{\pm} = \hat{a}_{\pm}^{\dagger} \hat{a}_{\pm}$, from which the SNR may be inferred which directly relates to error probability via the relation $P_{\mathrm{err}} = \mathrm{erfc}\left(  \sqrt{M
\mathrm{SNR} }\right)/2 $, where $\mathrm{erfc}$ is the complementary error function.

Despite its efforts in closing the gap between the performances of a physically realisable QI set-up and the theoretical ideal, the OPA and PC receivers have proven to be sub-optimal. This is primarily due to the fact that it demands an optimal measurement on pairs of modes constituting a mixed-state, physically done through Gaussian local operations with photon-number resolving measurements. These measurements belong to the class of LOCCs, known to be sub-optimal for such a mixed-state procedure~\cite{calsamiglia2010local} hence the associated receiver designs' sub-optimality follows naturally. 

\subsubsection{Sum-frequency generating receiver}

Ref.~\cite{FFSFG} showed that\cdel{, exploiting sum-frequency generation (SFG),} an improved receiver is capable of saturating QI's QCB \cadd{by exploiting the reverse process by which the signal and idler modes are generated in the first place: sum-frequency generation (SFG)}. Equipping the proposed receiver architecture\cadd{, based on SFG,} with a feed-forward (FF) mechanism was shown to \cadd{further }push its performance to the Helstrom bound in the limit of low signal brightness.

SFG is the inverse process to SPDC; the signal-idler photonic mode pairs are combined into a single photon with frequency $\omega_P = \omega_S + \omega_I$, the sum of the two individual frequencies. \cadd{This is possible, albeit with very low probability, if the signal-idler pair illuminates a crystal identical to the one responsible for its generation.} 

The underlying idea is motivated by the fact that one can \cdel{choose}\cadd{construct} a unitary $\hat{U}$ \cadd{which} make\cadd{s} one of the POVMs for optimal binary discrimination of two equally likely states, $\hat{U}\rho_{0}\hat{U}^{\dagger}$ and $\hat{U}\rho_{1}\hat{U}^{\dagger}$, simply a projection onto the vacuum. \cadd{In light of this, }SFG ultimately takes\cadd{,} as input\cadd{,} low-brightness signal-idler mode pairs with either zero or non-zero cross correlations, and outputs a single mode which is either a vacuum or non-vacuum, respectively. After multiple SFG cycles and FF, this signal is subjected to photon counting measurements \cadd{capable of achieving the minimum error probability defined by the Helstrom bound}.

This receiver design was later used in the determination of QI's receiver operating characteristic (ROC) which gives the detection probability as a function of the false alarm probability~\cite{QuntaoOSA}, in the Neyman-Pearson setting of QI~\cite{WTLB17}.

Nevertheless, despite its theoretical capability of saturating the QI QCB, such a receiver remains physically out of reach due to limitations on current technology, requiring both single-photon nonlinearities as well as efficient idler storage.

\subsubsection{Correlation-to-displacement receiver}

A more recent architecture for obtaining QI's full 6 dB quantum advantage, described in Ref.~\cite{shi2023fulfilling}, is the correlation-to-displacement receiver. Circumventing some of the practical limitations of the SFG receiver, a conversion module was proposed which worked by capturing the crucial phase-sensitive cross-correlations between $M$ signal-idler mode pairs and converting them to a coherent quadrature displacement. Essentially, the problem of two-mode detection, as in QI, is mapped onto a relatively straightforward, and more feasible, single-mode detection  problem of a noisy coherent state. The module itself requires the use of an $M \times M$ programmable beam splitter whose parameters are conditioned on the heterodyne measurement results of each of the returning modes $\hat{a}_{R}^{(k)}$. Note that the device still requires a reliable quantum memory for idler storage and, also, that one of the fundamental requirements for quantum target detection is that $M \gg 1$. This makes the implementation of such a device quite a formidable task, particularly for operation at microwave frequencies.

\cadd{The correlation-to-displacement architecture was applied to microwave QI in Ref.~\cite{angeletti2023microwave} where it was extended to accommodate experimental inefficiencies associated with this regime. By introducing a loss to to the return mode, associated with non-ideal coupling with the receiver, results showed that even with amplification a good quantum advantage compared to the ideal classical system can be guaranteed. This amounts to the full 6dB advantage in error exponent when the amplification is quantum-limited, reducing to 3dB when the amplifier introduces room temperature noise.}

\subsubsection{Hetero-homodyne receiver with sequential detection}

The latest proposed architecture for QI receiver is presented in Ref.~\cite{reichert2023quantum}. The novel design takes inspiration from the previously described correlation-to-displacement conversion~\cite{shi2023fulfilling} alongside sequential detection employed in Ref.~\cite{shapiro2022first} to beat Nair's pure-loss quantum radar performance limit~\cite{nair2011discriminating}. The proposal was a hetero-homodyne receiver, a cascaded POVM, which bypasses all previous requirements for a \cadd{joint} quantum measurement between the returning signal and the stored idler modes. \cadd{Returning signal modes are first heterodyne detected and the quadratures are used to inform the modulation of a microwave local oscillator, a coherent state field whose mean field is dependent on the heterodyne measurement outcome. This field is then used in an ideal homodyne detection of the stored idler modes. On its own, the hetero-homodyne receiver is shown to achieve the same 3dB performance enhancement over optimal coherent state illumniation as with Guha and Erkmen's~\cite{guha2009gaussian} PC and OPA receivers. It does so with the additional benefit of not requiring any sort of quantum interaction between the returning signal and idler beams.} 

\cdel{Alongside}\cadd{Used alongside} sequential detection, it was shown that the hetero-homodyne receiver offers a 6 dB enhancement in target detection capabilities for both classical (coherent state) and quantum illumination schemes, amounting to, for QI, a total quantum advantage of 9 dB over conventional classical radar. In essence, sequential detection provides an additional 3 dB in effective SNR per pulse. Thus, it is expected that the addition \cadd{of} sequential detection can further push the performances of the other known receivers capable of saturating the QI Chernoff bound\cdel{:} \cadd{such as} th\cdel{ose}\cadd{at} based on FF-SFG\cdel{ and correlation-to-displacement}. The trade-off for this is an increasingly complex receiver design on top of the base design which remains, so far, experimentally out of reach. It is also worth noting here that the hetero-homodyne receiver still assumes perfect idler preservation for the full range delay of the region of interest (equal to $2 R/c$ for a range $R$) and is still subject to the single-bin interrogation limit faced by standard QI. \cadd{The latter issue is exacerbated in sequential detection schemes since the interrogation time is necessarily longer compared to non-sequential, single pulse detection. As such, applying sequential detection to scenarios involving a moving target would prove problematic.}

\subsection{Quantum target ranging}\label{subsec:targetranging}

Classical radars are clearly capable of carrying out tasks outside of the simple detection of a target at some fixed distance. Through measurement of arrival times, a target's range may be inferred and, through detection of Doppler shift and the way a target's motion shifts the frequency of the returning signal, its velocity may be determined. The extension of QI from a problem of simple on/off detection to one of a realistic radar-like measurement is still an open question. Among various other issues, limitations are also due to the issue with idler storage mentioned in Sec.~\ref{sec:idlerstorage}. When the range of the target is unknown, the correct time for successful signal-idler recombination is also unknown. Without such knowledge, for the two-particle QI protocol, any joint measurement, be it pairwise or global (optimal) is impossible. For this and other reasons, quantum target ranging is a difficult task~\cite{karsa2020energetic}.

Ref.~\cite{maccone2020quantum} proposed a version of entanglement-based quantum radar to enable the localisation of a point-like target in three-dimensional space. In its most general form, the signal is a multipartite quantum state comprising $N$ entangled photons. All $N$ photons are used to probe the target region (there is no idler present in the scheme) and their relative positions and arrival times at the receiver may be used to infer the location of the target through electromagnetic scattering relations. Such an $N$-partite entangled probe is able to achieve a $\sqrt{N}$ decrease in uncertainty of target's position. As noted by the authors, there are two main issues associated with this protocol: firstly, it is notoriously difficult to generate such $N$-partite maximally entangled states for large enough $N$ to yield any noticeable benefit. While the generation of $N=2$ entangled states may be done through SPDC optical sources, the loss of just one of the $N$ modes renders the remaining $N-1$ modes completely useless. Secondly, the model assumes total randomness in returning signal arrival times and transverse locations on a planar receiver. In other words, it requires an infinite measurement time and receiver size in order to obtain these results. The use of a more complex, nested system of entangled state could address this problem while non-maximally entangled states could address both due to their increased noise resilience. In either case, there is a corresponding decrease in potential enhancement of measurement precision

\subsection{Issue with loss tolerance}

QI is robust to noise but not to loss. It provides advantage in the task of distinguishing an entangled signal photon within a bright thermal background, but as long as the overall round-trip loss is limited~\cite{lanzagorta2012quantum}. While this loss may be limited in optical implementations with collimated Gaussian beams, it fails to be the case in the microwave regime where the typical wave is much more isotropic and therefore subject to high geometric loss\cdel{ (already in the forward path)}. \cadd{As a result, even in the forward path, the outgoing beam is subject to increased losses which are easily alleviated at shorter wavelengths.} 

One can mitigate this geometric loss but this requires high gain antennas (large arrays). \cadd{Other techniques have also been proposed which involve the use of adaptive correction~\cite{smith2009correction} however practically doing this requires prior knowledge of several parameters governing the overall attenuation, including the target range. Clearly this is problematic if one wishes to achieve a quantum radar, though the techniques could be beneficial for, for example, fixed-ranged quantum scanning applications.}

\subsection{Issue with detection time}

Typical QI advantage is provided in the exponent of the mean error probability under the assumption of many signals emitted, i.e., for $M \gg 1$. However, this means an overall long detection time if the \cdel{clock}\cadd{detection rate} of the signals is not sufficiently high. For example, consider a carrier frequency of $1$GHz (L band). If we assume broadband signals with $10\%$ bandwidth ($100$MHz), their time duration is $10$ns. Assuming $M = 10^8$ signals as typical in QI works, one is faced with an overall integration time of about 1 second. This is clearly too slow for standard fast-moving targets. However, this may be acceptable for very short-range surveillance of slow objects, or even stationary ones prevalent within biomedical applications.

\section{Prototypical experiments on QI}\label{sec:QIexperiments}

The first QI experiment in the optical domain was carried out in Ref.~\cite{lopaevaexp} using an SPDC source and photon counting to detect a target modelled by a 50:50 beam splitter. They demonstrated that a QI-like advantage in effective SNR compared to a correlated thermal state may be achieved in a thermal background when the channel used was entanglement-breaking. The chosen classical benchmark was not the ideal one based on coherent light. (The same is true for a similar, more recent, experiment~\cite{england2019quantum}.) Later on, Refs.~\cite{zhangent,zhangexp} implemented the optical Gaussian QI protocol using an OPA receiver. Their experiment demonstrated a sub-optimal 20\% improvement (equivalent to 0.8 dB in comparison to the 3 dB available with OPA receivers) with respect to the optimal performance of coherent stares (ideal benchmark).

After the theoretical work on microwave QI~\cite{barzanjeh2015microwave}, there have been experimental demonstrations at these longer wavelengths~\cite{chang2019quantum,luong2019receiver,shabirQI}. All three experiments employed a Josephson parametric converter (JPC) for entanglement generation of microwave modes with low-brightness. After JPC generation, both modes were amplified and the signal was sent to a room-temperature target region while the idler was immediately heterodyne detected. The classical outcome of this heterodyne detection was stored digitally to be compared with the outcome of the heterodyne detected returning signal in post-processing. In all experiments, a relative QI advantage was discussed over their chosen classical comparison cases (such as the generation of microwave coherent states via the same setup). No quantum advantage was shown with respect to the ideal classical benchmark.

More recently, Ref.~\cite{Huard2022} performed a completely cryogenic experiment aiming to demonstrate a quantum advantage in microwave quantum radar. They used a superconducting circuit kept at 15 mK to first generate microwave TMSV quantum states and then store the idler mode while allowing the signal to travel down a delay line coupled with a tunable filter to model varying target reflectivity. Thermal noise is then injected into the reflected probe signal before the receiver. The final step, measurement, comprises a similar squeezing operation to the TMSV generation technique effectively implementing the OPA receiver of Guha and Erkmen~\cite{guha2009gaussian} with some gain $G$ which can, ideally, be optimised to recombine the reflected signal and idler. Then, the quantum correlations may be accessed by simply measuring the idler's average photon number. Their results displayed an average error exponent advantage of $\sim 0.8$ dB over the theoretical classical benchmark, which is on par with what has already been achieved in optics (see Refs.~\cite{zhangent,zhangexp}).

As mentioned earlier, none of these experiments can be regarded as a realistic implementation\cdel{s} of quantum radar. They are, however, important demonstrations of the potentialities of QI and, more generally, quantum target detection under restrictive assumptions enabling us to examine some of the underlying principles of QI, particularly when operating within the microwave domain.

\section{Further modifications of the QI protocol}\label{sec:extensions}

There have been multiple attempts at increasing sensitivity in target detection by increasing the number of modes in which entanglement is distributed, as already discussed in Sec.~\ref{subsec:targetranging} for improved sensitivity in target localisation. The improvement in sensitivity for Gaussian QI using three entangled modes, where one is labelled the idler while the other two are sent as signal beams, has been shown in Refs.~\cite{gallego2021quantum}. It has also more recently been applied to the case of non-Gaussian three-mode entangled states~\cite{torrome2023non}. The increase in degree of entanglement (in terms of the number of modes) is shown to provide this broader form of QI more robust to noise, as well as potentially loss. A similar idea has also been studied in Ref.~\cite{jung2022quantum} with three-mode Gaussian entanglement involving, this time, one signal and two idler beams. Their analysis of error bounds found an improvement compared to the two-mode Gaussian QI protocol when the average number of photons per mode in the signal $N_S < 0.295$. While these are promising results, the generation of such three-photon entangled states is a non-trivial task, as discussed in Sec.~\ref{subsec:targetranging}, though state generation of this type remains an active field of research.

There exists wide interest across the quantum technology community in the study of non-Gaussianity and, with respect to QI, how it may be introduced to enhance the existing protocol. While Gaussian quantum states, operations and detection schemes offer a relatively straightforward means of physical implementation and mathematical description, their use together often places limitations within the realms of potential quantum enhancements. Alternative approaches of making use of non-Gaussianity include, for example, probe states where non-Gaussianity is introduced via photon addition or photon subtraction~\cite{zhang2014quantum,fan2018quantum}. It has also been shown that non-Gaussian receiver designs, in the form of a non-linear amplifier (NLA), has the potential of enhancing the effective SNR of QI to the extent of extending the range of possible regimes for which a quantum advantage over coherent states can be obtained~\cite{karsa2022NLA}.

\section{Conclusion}\label{sec: outlook}

Despite many advances in QI, the possibility of a quantum radar still seems to be 
far in the future. There are a number of theoretical and experimental challenges to overcome
which impede the fulfillment of the essential criteria (see Sec.~\ref{sec:criteria}). In particular,
Type-3 designs, based on entanglement distribution (like QI), appear to be too fragile to loss in order 
to be implemented in a practical microwave scenario. For this and other reasons, it is extremely unlikely to foresee any near-term development of a long-range quantum radar. Besides the distance limitations, there are the challenges associated with the actual ranging capabilities and detection times (as discussed \cdel{in detail}in Sec.~\ref{sec:exp reqs}). Because the possibility of a long-range prototype is remote, currently there is low chance of potential military use of QI-based technology.

That being said, applications of QI appear to be feasible for short-range detection of fixed or slowly-moving
objects. For example, one could consider the quantum-enhanced scanning of metallic objects in a surveillance scenario. The ranging issue is not important in this case, since the main question is the presence or not of a reflection, while slowly changing the depth of the scan (so the round-trip distance would be fixed for the probing of each layer). It is also clear that the interrogation time can be much longer than what required in radar-like detection. The advantage of the quantum design relies in lower error probabilities while reducing the mean number of photons required for the scan. The latter property of small irradiated energy can also be associated with biomedical settings. 

\section*{Acknowledgments}
This work was funded by the EU Horizon 2020 Research and Innovation Action under Grant Agreement No.~862644 (FET Open project: Quantum read-out techniques and technologies, QUARTET). 

\end{document}